\documentclass{article}
\usepackage{slashed}
\usepackage[dvips]{graphicx}

\newcommand{\lsim}{\raisebox{-4pt}{$\,\stackrel{\textstyle
                                                         <}{\sim}\,$}}

\newcommand{\nn}{\nonumber}
\newcommand{\be}{\begin{equation}}
\newcommand{\ee}{\end{equation}}
\newcommand{\ba}{\begin{eqnarray}}
\newcommand{\ea}{\end{eqnarray}}
\newcommand{\req}[1]{(\ref{#1})}
\def\={\,=\,}
\newcommand{\ci}[1]{\cite{#1}}

\def\mev{~{\rm MeV}}
\def\gev{~{\rm GeV}}

\def\ale{\alpha_{\rm em}}
\def\als{\alpha_{\rm s}}

\def\taub{\bar{\tau}}
\newcommand{\tw}{\textwidth}

\def\vb0{{\bf b}_0}

\def\={\,=\,}

\begin{document} 
\thispagestyle{empty}
\begin{flushright}
January, 31  2019\\[20mm]
\end{flushright}

\begin{center}
{\Large\bf 
Hard exclusive processes involving kaons}\\
\vskip 10mm
P.\ Kroll  \footnote{Email:  kroll@physik.uni-wuppertal.de} \\[1em]
{\small {\it Fachbereich Physik, Universit\"at Wuppertal, D-42097 Wuppertal, Germany}}\\ 
\end{center}
\vskip 5mm 
\begin{abstract}
\noindent 
Hard exclusive electroproduction of kaons as well as the kaon-induced exclusive Drell-Yan
process are investigated within the handbag approach which is based on factorization in
hard subprocesses and soft generalized parton distributions (GPDs). The kaon-hyperon 
transition GPDs occurring here, are related to the proton GPDs by flavor symmetry. The 
latter ones are taken from analyses of pion electroproduction. Like in hard processes 
involving pions the transversity GPDs play an important role in the processes of interest 
- the transverse cross sections are larger than (or, for the Drell-Yan process,  about 
equal to) the longitudinal ones. The evolution of the transversity GPDs is taken into 
account for the first time but, as the analysis reveals, it is a minor effect in the range 
of photon virtualities of interest. The predictions for the cross sections agree fairly 
well with the sparse available electroproduction data. 
\end{abstract}

\section{Introduction}
\label{eq:intro}
The handbag approach to hard exclusive meson electroproduction is based on 
factorization of the process amplitudes in hard subprocesses and soft hadronic
matrix elements, parametrized as GPDs. This factorization property has been shown 
to hold rigorously to leading-twist accuracy in the generalized Bjorken regime  of 
large photon virtuality, $Q$, and large invariant mass of the hadrons in the final 
state, $W$, but fixed Bjorken-x,
$x_B$, and small Mandelstam-$t$ \ci{collins97}. From extensive experimental and 
theoretical investigations of hard exclusive meson electroproduction carried through 
over the last two decades it however turned out that the naive asymptotic result 
is not readily applicable in the range of kinematics accessible to current experiments.
In fact, large power corrections are required to the asymptotically dominant 
amplitudes for longitudinally polarized photons. Moreover there are strong 
contributions from transversal photons which are asymptotically suppressed by $1/Q^2$ 
in the cross sections. In some cases, as for instance for $\pi^0$ electroproduction 
\ci{dufurne}, the contributions from transversely polarized photons are even dominant. 

In a series of articles we have developed a generalization of the handbag approach 
which allows to model these power corrections, see for instance the detailed report 
\ci{kroll14}. The decisive point is to retain the  quark transverse momenta in the 
subprocess. Implicitly, this way the transverse size of the meson is taken into account. This 
generalized handbag approach has been applied to electroproduction 
of pions \ci{GK5,GK6} as well as to $\rho^0$ and $\phi$ mesons \ci{GK3}. It turned 
out that the data on these processes are well fitted within this approach in a large 
range of kinematics. An outcome of these investigations is the extraction of
a set of GPDs which subsequently allow to study the parton localization in
the transverse position plane, to evaluate the parton angular momentum and,
exploiting the universality properties of the GPDs, to calculate other hard exclusive
processes as for instance deeply virtual Compton scattering \ci{KMS12} or $\omega$
production \ci{GK8}. 

In this article I am going to apply the generalized handbag approach to hard exclusive
processes involving kaons. Not much has been done as yet for these processes neither
theoretically nor experimentally. Only a few data on the separated electroproduction 
cross sections for forward emitted kaons have been measured at the Jefferson lab 
\ci{mohring,coman,carmignotto}. More data will come from the JLab experiment E12-09-011
in the near future. The kaon-induced exclusive Drell-Yan process 
is planned to measure at J-Parc  \ci{chang}. Thus, it seems to be of interest
and timely to study these kaon reactions in order to probe the set of extracted GPDs 
against kaon data and to make predictions for future experiments.

The plan of the paper is the following: In the next section the generalized handbag 
approach is briefly sketched and the soft input (GPDs, kaon wave functions, kaon-pole
term) is represented. Results for kaon electroproduction are given in Sect. 3 and
compared to the data. In Sect. 4 predictions for the Drell-Yan process are presented
and the implications of the excitation of charmonia are examined.
A summary is given in Sect. 5 and in the appendix a method for the numerical solution 
of the evolution equation for the transversity GPDs is discussed.
\section{The handbag approach}
\label{sec:handbag}
The generalized handbag aproach has been described in great detail in previous work
\ci{GK5,GK6,GK3}. Therefore, only the basics facts will be sketched here.
Consider the process $\gamma^*(q,\mu) p(p,\nu) \to K^+(q',0) \Lambda(p',\nu')$
in the generalized Bjorken-regime. The symbols in the brackets denote the momenta 
and helicities of the respective particles. Mandelstam-$t$ is assumed to be much 
smaller than $Q^2$. Therefore, terms of order $(\sqrt{-t}/Q^2)^n$, $n\geq 2$
are neglected throughout. The helicity amplitudes for electroproduction of kaons 
are given by convolutions of subprocess amplitudes and suitable flavor combinations 
of GPDs. The amplitudes read \ci{GK6}
\ba
{\cal M}_{0+,0+}&=& \sqrt{1-\xi^2}\frac{e_0}{Q}\,\left[<\widetilde H_K>_0 
                  -\frac{\xi^2}{1-\xi^2}<\tilde E_K^{\rm n.p.}>_0 \right]\,,\nn\\
{\cal M}_{0-,0+}&=& \frac{e_0}{Q}\,\frac{\sqrt{-t+t_0}}{m+m_\Lambda}\,\xi\,
                                               <\tilde E_K^{\rm n.p.}>_0\,, \nn\\
{\cal M}_{0-,++}&=& \frac{e_0\mu_K}{Q^2}\sqrt{1-\xi^2}\,<H_{TK}>_+ \,, \nn\\
{\cal M}_{0+,\pm +}&=& \frac{e_0\mu_K}{Q^2}\,\frac{\sqrt{-t+t_0}}{m+m_\Lambda}\,
                                               <\bar{E}_{TK}>_+\,.
\label{eq:amplitudes}
\ea
The amplitude ${\cal M}_{0-,-+}$ is neglected since it is suppressed by $(t-t_0)/Q^2$.
Explicit helicities are labeled by their signs only or by zero. The proton and the 
$\Lambda$ masses are denoted by $m$ and $m_{\Lambda}$, respectively. The meson mass, 
$m_K$, is neglected except in the meson-pole term, see Sect.\ \ref{subsec:pole}. The 
amplitudes for negative helicity of the initial state proton are obtained from the 
above set of amplitudes by parity conservation. The positron charge is denoted by 
$e_0$ and the skewness, $\xi$, is related to Bjorken-$x$, $x_B$, by
\be
\xi\=\frac{x_B}{2-x_B}
\label{eq:skewness}
\ee
where possible corrections of order $1/Q^2$ are ignored. The minimal value of $-t$ 
corresponding to forward scattering, is given by
\be
t_0\=-\frac{2\xi}{1-\xi^2}\,\Big[m_{\Lambda}^2\,(1+\xi) - m^2\,(1-\xi)\Big]\,.
\label{eq:t0}
\ee

The characteristic $Q$-dependencies of the amplitudes have been made explicitly in
Eq.\ \req{eq:amplitudes}. For dimensional  reasons the mass parameter, $\mu_K$, is 
additionally pulled out from the convolutions, $<K_K>$, for transversely polarized photons. 
This parameter is the meson mass enhanced by the chiral condensate
\be
\mu_K\=\frac{m_K^2}{m_u+m_s}
\label{eq:mass-parameter}
\ee
by means of the divergence of the axial vector current. The masses $m_u$ and $m_s$ are
the current-quark masses of the kaon's valence quarks. For the numerical studies to be 
presented below, a value of $2\,\gev$ is used for $\mu_K$ at the initial scale $\mu_0=2\,\gev$. 
Since the current-quark masses decrease with increasing scale $\mu_K$ is scale dependent. 
The respective anomalous dimension is $4/\beta_0=12/25$ for four flavors. The mass parameter, 
$\mu_K$, occurs since the use of the transversity or helicity-flip GPDs, $H_T$ and $\bar{E}_T$,  
goes along with the twist-3 kaon wave function which is applied in Wandzura-Wilczek 
approximation. As one sees from Eq.\ \req{eq:amplitudes} the transverse amplitudes are 
parametrically suppressed by $\mu_K/Q$ as compared to the longitudinal amplitudes which 
are of twist-2 nature.

The item $<K_K>$ in \req{eq:amplitudes} denotes the convolution of a proton-hyperon 
transition GPD, $K$, with a subprocess amplitude, ${\cal H}$:
\be
<K_K>_\mu\=\sum_{\lambda'\lambda} \int_{-1}^1dx {\cal H}_{0\lambda',\mu\lambda}^K(x,\xi,Q^2,t=0)
                                             \,K_K(x,\xi,t)\,.
\ee
The labels $\lambda$ and $\lambda'$ refer to the helicities of the partons participating 
in the subprocess. The subprocess amplitudes are calculated with the quark transverse 
momenta retained in the subprocess while the emission and reabsorption of the partons 
from the baryons are still treated collinear to the baryon momenta. The subprocess 
amplitudes read
\ba
{\cal H}_{0\lambda',\mu\lambda} &=& \int d\tau d^2b\, \hat{\Psi}_{K,-\lambda'\lambda}
                                                              (\tau,-{\bf b},\mu_F)\,
                      \hat{F}^K_{0\lambda'\mu\lambda}(x,\xi,\tau,Q^2,{\bf b},\mu_R)\nn\\
                  &&\times           \als(\mu_R)\exp{[-S(\tau,{\bf b},Q^2,\mu_F,\mu_R]}
\label{eq:subprocess}
\ea
in the impact parameter space; ${\bf b}$ is canonically conjugated to the quark transverse
momenta. The Sudakov factor, $S$, has been calculated by Botts and Sterman \ci{botts} in
next-to-leading-log approximation using resummation techniques and having recourse to the
remormalization group. It takes into account the gluon radiation resulting from the 
separation of color charges which is a consequence of the quark transverse momenta.
The Sudakov factor can be found in \ci{kroll10}. Its properties force the following
choice of the factorization scale: $\mu_F=1/b$. The renormalization scale is taken to be
the largest mass scale appearing in the subprocess, i.e. 
$\mu_R={\rm max}(\tau Q,(1-\tau) Q,1/b)$ ($\tau$ is the momentum fraction of the quark 
entering the meson). The renormalization scale also applies to the mass parameter
defined in Eq.\ \req{eq:mass-parameter}.
For the hard scattering kernel, $F$, evaluated to lowest order of 
perturbative QCD, it is referred to Refs.\ \ci{GK5,GK3}. The last item to be explained 
is $\hat{\Psi}_{K,\lambda'\lambda}$. It represents the Fourier transform of a meson 
light-cone wave function. For longitudinally polarized photons, one has 
$\lambda'=\lambda$ and the distribution amplitude associated to $\hat{\Psi}$, is the 
familiar twist-2 one. For transversal 
photons one has $\lambda'=-\lambda$ and a twist-3 wave function is required. The wave 
functions are specified in Sect.\ \ref{subsec:wf}. The inclusion of the quark transverse 
momenta and the Sudakov factor has two advantages - firstly the magnitude of the subprocess 
amplitudes are somewhat reduced as compared to a collinear calculation which leads to a 
better agreement with experiment (see e.g.\ \ci{GK6}) and, secondly, the infrared 
singularities occurring in a collinear calculation of the twist-3 subprocess amplitudes are 
regularized. In passing it should be noted that there are other $1/Q$ suppressed 
contributions to the transverse amplitudes, as for instance twist-3 GPDs in combination 
with leading-twist meson wave functions. Since for such contributions there is no 
enhancement known it seems reasonable to neglect them and to include just the combined 
effect of the transversity GPDs and the twist-3 meson wave functions.

\subsection{The GPDs}
\label{subsec:GPD}

For the process of interest in this work the proton-$\Lambda$ transition GPDs, $K^K$,
occur. Flavor symmetry however relates these GPDs to the diagonal proton ones
\ci{frankfurt99}. On the premise of a flavor symmetric sea only valence quarks contribute
to kaon electroproction and the  proton-$\Lambda$ GPDs are given by
\be
K_{Ki}\simeq - \frac1{\sqrt{6}}\Big[ 2 K_i^{u} - K_i^{d}\big]\,.
\ee
In \ci{GK6} the zero-skewness GPDs for flavor $a$ are parametrized as
\be
K_i^a(z,\xi=0,t) \= K_i^a(x,\xi=t=0) \exp{[(b_i^a-\alpha_i'{}^a\ln(x))t]}\,.
\label{eq:profile}
\ee
The forward limits, $\xi,t\to 0$, of the GPDs $\widetilde{H}$ and $H_T$ are given by 
the polarized and transversity parton densities, respectively. In order to respect
the Soffer bound the transversity density is parametrized as \ci{anselmino}
\be
\delta^a\=N^a_{H_T} \sqrt{x}\,(1-x)\,\Big[q^a(x)+\Delta q^a(x)\Big]\,.
\ee
The unpolarized and polarized parton densities are taken from \ci{ABM11} and \ci{DSSV09},
respectively. For the $E$-type GPDs the forward limits are not accessible in deep ineleastic 
lepton-nucleon scattering and, hence, unknown. Therefore, they are parametrized like the PDFs
\be
K_i^a(x,\xi=t=0)\= N_i^a x^{-\alpha_i^a(0)}(1-x)^{\beta_i^a}
\ee
with the additional parameters to be adjusted to the electroproduction data. The 
products of the zero-skewness GPDs with suitable weight functions are considered as 
double distributions from which the full GPDs can be calculated \ci{radyushkin}. The 
parameters of the GPDs, compiled in Tab.\ \ref{tab:gpd}, are taken from \ci{GK6}. 
The powers $\beta_i^a$ are set to the following values
\ba
\tilde{E}^{\rm n.p.}:&&  \qquad  \beta^u=5\,, \qquad  \beta^d=5\,,  \nn\\
\bar{E}_T: && \qquad \beta^u=4\,, \qquad  \beta^d=5\,.
\ea
It should be noticed that two variants of the transversity GPDs are discussed
in \ci{GK6} . One which leads to a rather deep dip in the $\pi^0$ cross section for 
forward scattering while for the second one the dip is less deep. Since the second 
one is in better agreement with the $\pi^0$ electroproduction data \ci{dufurne,CLAS} 
it is used here.

\begin{table*}[t]
\renewcommand{\arraystretch}{1.4} 
\begin{center}
\begin{tabular}{| c || c | c | c || c | c |}
\hline   
GPD & $\alpha(0)$ & $\alpha^\prime [\gev^{-2}]$ & $b [\gev^{-2}]$ & $N^u$ &
$N^d$ \\[0.2em]  
\hline
$\widetilde{H}$ & - & 0.45  &  0.59  & -  & - \\[0.2em]
$\tilde{E}^{\rm n.p.}$ & 0.48 & 0.45 & 0.9 & 14.0 & 4.0 \\[0.2em]
$H_T$ & - & 0.45 & 0.3 & 1.1 & -0.3 \\[0.2em]
$\bar{E}_T$& 0.3 & 0.45 & 0.5 & 4.83 & 3.57 \\[0.2em]
\hline
\end{tabular}
\end{center}
\caption{Regge parameters and normalizations of the valence-quark GPDs, quoted at a scale
 of $\mu_0=2\,\gev$ for $\widetilde{H}$ and $\tilde{E}^{\rm n.p.}$ and at $1.41\,\gev$ for the 
transversity GPDs. The GPD $\tilde{E}^{\rm n.p.}$ represents $\tilde{E}$ with the kaon pole
subtracted. Lacking parameters indicate that the corresponding parameters
are part of the parton densities.}
\label{tab:gpd}
\renewcommand{\arraystretch}{1.0}   
\end{table*} 

\begin{figure}
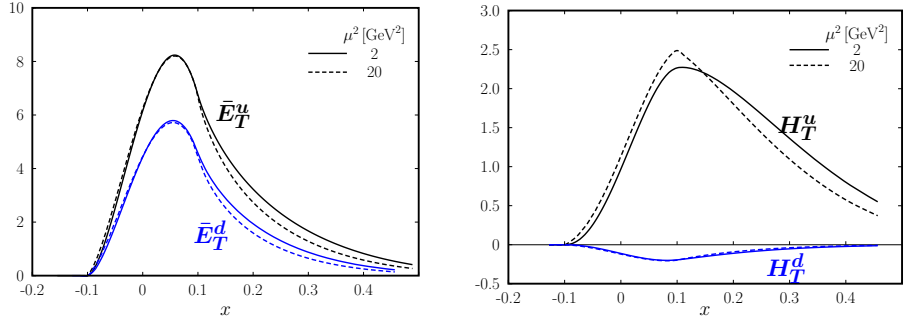

\centering
\includegraphics[width=0.45\tw]{fig-EbarTu-2-20.epsi}\hspace*{0.05\tw}
\includegraphics[width=0.47\tw]{fig-HTud-2-20.epsi}
\caption{The evolution of the proton transversity GPDs $\bar{E}_T$ (left) and $H_T$ (right)
from 2 to 20 $\gev^2$ at $t=-0.036\,\gev^2$ and $\xi=0.1$.}
\label{fig:trans_GPDs_evolv}
\end{figure}

As is well-known the GPDs evolve with the scale. The evolution of the helicity non-flip
GPDs, $\widetilde{H}, \tilde{E}^{\rm n.p.}$, is evaluated with the help of Vinnikov's code 
\ci{vinnikov}. In contrast to previous applications of the transversity GPDs \ci{GK5,GK6,liuti} 
their scale dependence is also taken into account in this work. The numerical method used 
to compute the evolution is described in the appendix. It turns out that the evolution of 
the transversity GPDs is a minor effect within the range of scales accessible to current 
electroproduction experiments as can be seen from Fig.\ \ref{fig:trans_GPDs_evolv} where 
$\bar{E}_T$ and $H_T$ are shown at the scales $\mu^2=2$ and $20\,\gev^2$. That the evolution 
is a small effect is already signaled by the anomalous dimension of the lowest moment of 
$\bar{E}_T$ at $\xi=t=0$, the tensor anomalous magnetic moment of the nucleon,
\be
\kappa^a_T(\mu^2) \= \int_{-1}^1 dx \bar{E}^a_T(x,\xi=t=0,\mu^2)
\ee
which evolves as \ci{hoodbhoy,bel-mul}
\be
\kappa^a_T(\mu^2) \=\left(\frac{\als(\mu^2)}{\als(\mu_0^2)}\right)^{\gamma_0^T/\beta_0} 
                      \kappa^a_T(\mu_0^2)
\ee
with the anomalous dimension~\footnote{
     Anomalous dimensions are quoted for four flavors, $n_f=4$, throughout the paper.} 
\be
   \gamma_0^T/\beta_0\=C_F/\beta_0\=4/25\,.
\ee
The same scale dependence exhibits the tensor charge, the lowest moment of $H_T$ at
$\xi=t=0$.
As an example for the significance of the evolution of the transversity GPDs it is noticed
that the transverse cross section for $K^-p\to \gamma^*\Lambda$ which will be discussed in
Sect. \ref{sec:drellyan}, is reduced by mere $8\%$ at a photon virtuality of $14\,\gev^2$
if the evolution is taken into account.
Hence, the neglect of the evolution of the transversity GPDs in previous work \ci{GK5,GK6}
is justified.

Since the factorization of the GPDs and the subprocess is treated collinearly, the GPDs
do not know of the impact-parameter dependence in the subprocess - $b$ is integrated over.
Hence, the factorization scale $\mu_F$ does not apply to the GPDs, it refers to the factorization
of the soft meson wave function and the remaining hard part of the subprocess. The scale
of the GPDs is therefore taken as the photon virtuality.
\subsection{The meson wave functions}
\label{subsec:wf}
In contrast to the GPDs which are universal, i.e.\ process independent, the subprocess
amplitudes depend on the meson by means of the meson wave function.
For the soft twist-2 kaon light-cone wave function the following form is used
\be
\Psi_{K,-+}\=8\pi^2 \frac{f_K}{\sqrt{2N_c}}\,\frac{\zeta_K^2}{\tau\taub}\,\Phi_K(\tau)
                   \exp{[-\zeta_K^2k^2_\perp/(\tau\taub)]}
\label{eq:twist-2-wf}
\ee
with the distribution amplitude ($\taub=1-\tau$)
\be
\Phi_K(\tau)\=6\tau\taub\,\Big[1 + a_{K1}\,C_1^{3/2}(2\tau-1) 
                                  + a_{K2}\,C_2^{3/2}(2\tau-1) + \ldots\Big]\,.
\label{eq:DA}
\ee
For the transverse size parameter of the kaon the same value as for the pion is taken
\be 
\zeta_K\=\zeta_\pi\=0.853\,\gev^{-1}\,.
\ee
The kaon decay constant is $159\,\mev$ \ci{PDG}.
The Gegenbauer coefficients, $a_{Kn}$ evolve with the scale
\be
a_{Kn}(\mu_F^2)\=\left(\frac{\als(\mu_F^2)}{\als(\mu_0^2)}\right)^{\gamma_n/\beta_0} 
                                                                   a_{Kn}(\mu_0^2)
\ee
with the anomalous dimensions $\gamma_1/\beta_0=32/75$ and $\gamma_2/\beta_0=2/3$.
The first two Gegenbauer coefficients in \req{eq:DA} are assumed to take the values
\be
a_{K1}(\mu_0)\=0.086\pm 0.04\,,  \qquad a_{K2}(\mu_0)\=-0.159\pm 0.07\,.
\label{eq:coefficients}
\ee
All coefficients $a_{Kn}$ for $n\geq 3$  are set to zero. This is justified to some extent
because the Sudakov factor in conjunction with the hard scattering kernel suppresses the 
contributions from the higher Gegenbauer terms as compared to the lowest term at not too 
large values of the factorization scale \ci{kroll10}. The strength of the suppression grows 
with the Gegenbauer index. With the values \req{eq:coefficients} the antistrange quark in the
$K^+$ carries a smaller momentum fraction on the average than the $u$-quark. 
A negative value of $a_{K2}$ is chosen in order to avoid an overestimate of kaon electroproduction.
Kaon channels are typically suppressed by about $10\%$ as compared to pion channels. This, for
instance, can be seen in the time-like electromagnetic form factors \ci{cleo13}, in two-photon
annihilation \ci{belle04} or in $\chi_{cJ}$ decays into pairs of mesons \ci{PDG}. However, 
from QCD sum rules \ci{ball06} and the Dyson-Schwinger approach \ci{chang14} a positive Gegenbauer 
coefficient $a_{K2}$ has been found.

The twist-3 light-cone wave function is assumed to be
\be
\Psi_{K,++}\=\frac{16\pi^{3/2}}{\sqrt{2N_c}}\,f_K \zeta_{KP}^3k_\perp \exp{[-\zeta_{KP}^2k_\perp^2]}
\ee
with the associated pseudoscalar twist-3 distribution amplitude $\Phi_{KP}\equiv 1$ 
\ci{braun-filyanov}. The transverse size parameter $a_{KP}$ is set to the same value as in the 
pion case \ci{GK9}, namely $1.8\,\gev^{-1}$. It should be mentioned that there is a second 
two-body twist-3 wave function, the tensor one. It has been shown \ci{GK5}, however, that its 
contribution to the subprocess amplitude is proportional to $t/Q^2$ and, hence, neglected. 
Also neglected are possible contributions from the three-body twist-3 wave function. 

\subsection{The kaon-pole term}
\label{subsec:pole}

The kaon-pole contribution is treated as a one-boson-exchange contribution \ci{GK5}
which leads to the amplitudes:
\ba
{\cal M}_{0+,0+}^{\rm pole} &=& -\frac{e_0}{Q}\frac{(m+m_\Lambda)\xi}{\sqrt{1-\xi^2}}
                          \,\frac{\rho_K}{t-m_K^2}\,, \nn\\
{\cal M}_{0-,0+}^{\rm pole} &=& \frac{e_0}{Q} \sqrt{-t+t_0}\,\frac{\rho_K}{t-m_K^2}\,, \nn\\
{\cal M}_{0+,\pm +}^{\rm pole} &=&\pm \sqrt{2}\,\frac{e_0}{Q^2}\,\sqrt{-t+t_0}(m+m_\Lambda)\xi
                         \,\frac{\rho_K}{t-m_K^2}\,, \nn\\
{\cal M}_{0-,\mu +}^{\rm pole} &=&0\,.
\label{eq:pole}
\ea
These amplitudes have to be added to those in Eq.\ \req{eq:amplitudes}. The pole contribution
is free of evolution and not subject to higher-order perturbative QCD corrections.
The residue of the kaon pole is given by
\be
\rho_K\=\sqrt{2} g_{KN\Lambda} F_{KN\Lambda}(t) Q^2F_K^{\rm s.l.}(Q^2)\,.
\label{eq:residue}
\ee
The kaon-baryon coupling constant is taken as 
\be
g_{KN\Lambda}\=-14.5\pm 1.3\,.
\ee
This value is a combination of results  quoted in \ci{compilation} with a more recent
value extracted from data on $p\bar{p}\to\Lambda\bar{\Lambda}$ \ci{timmermans}. 
There is also a form factor for the coupling of the kaon to the baryons which is
parametrized as
\be
F_{Kp\Lambda}\= \frac{\Lambda_{p\Lambda}-m_K^2}{\Lambda_{p\Lambda}-t}
\ee
where $\Lambda_{p\Lambda}=1.1\pm 0.08\,\gev$. The last item in \req{eq:residue} to be specified
is the electromagnetic form factor of the kaon in the space-like region. It is parametrized as 
\be
Q^2F_K^{\rm s.l.}(Q^2)\= \big[1+Q^2/c_K^{\rm s.l.}\big]\,.
\label{eq:sl-FF}
\ee
The parameter $c_K^{\rm s.l.}$ is taken as $0.5\pm 0.04\,\gev^2$ in agreement with JLab data 
\ci{carmignotto}.

To leading-twist accuracy the kaon pole can be viewed as part of the GPD $\tilde{E}$
\ci{penttinen99}. The convolution of this GPD with a hard subprocess amplitude lead
to the same longitudinal amplitudes as in \req{eq:pole} except that the electromagnetic 
form factor of the kaon is the leading-order perturbative result. This leads
to an underestimate of the pole contribution to the kaon electroproduction cross section.

The time-like kaon electromagnetic form factor which will be needed for the evaluation
of the kaon-induced exclusive Drell-Yan process, is taken from the CLEO data \ci{cleo13}. 
It is parametrized as 
\be
Q'{}^2 |F_K^{\rm t.l.}| \= c^{\rm t.l.}_K
\label{eq:tl-FF}
\ee
with 
\be
c^{\rm t.l.}_K\=0.80\pm 0.04\,.
\ee
Its phase factor, ${\rm exp}{[i\eta(Q'{}^2)]}$, is taken to be the same as for the time-like pion 
form factor for simplicity. For the latter phase the dispersion relation result of \ci{dubnik11} 
is used. According to this work, the phase, $\eta$, is close to the asymptotic phase $\pi$ for 
$Q'{}^2\lsim 8.9\,\gev^2$; exactly $\pi$ is taken for larger $Q'{}^2$. The asymptotic phase follows 
from perturbative QCD. 

\section{Kaon electroproduction}
\label{sec:electro}
\begin{figure}
\centering
\includegraphics[width=0.57\tw]{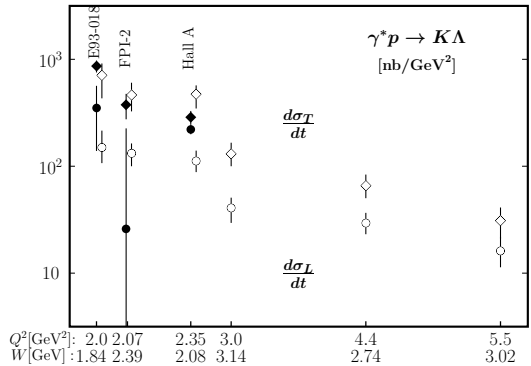}
\caption{The longitudinal and transversal cross sections for $\gamma^* p\to K^+\Lambda$.
The experimental data \ci{mohring,carmignotto,coman} are displayed by filled symbols, 
the theoretical results by open ones. Diamonds (circles) represent the transverse 
(longitudinal) cross sections. Data and predictions are at the respective $t_0$
except for $W=2.39\,\gev$, $Q^2=2.07\,\gev^2$ where $t=-0.4\,\gev^2$.}
\label{fig:kaon}
\end{figure}

With the universal GPDs at disposal and the information about the kaon, specified
in Secs.\ \ref{subsec:wf} and \ref{subsec:pole}, the partial cross sections for 
electroproduction of kaons can be computed. The results on the longitudinal and 
transverse cross sections for forward going kaons are shown in Fig.\ \ref{fig:kaon}
and compared to the available data \ci{mohring,carmignotto,coman}. Predictions for 
the kinematics chosen for the Jlab E12-09-011 experiment are also displayed in this
figure. Fair agreement with experiment is to be seen. The parametric uncertainties of
the predicted cross sections have been estimated from the errors of
the various parameters discussed in Sects.\ \ref{subsec:wf} and \ref{subsec:pole} as 
well as from an estimate of the uncertainties of the GPDs \ci{GK6,GK3,GK9}.
The evolution of the GPDs, the kaon distribution amplitude as well as that of the mass
parameter $\mu_K$ are taken into account. 
A remarkable fact is that even for forward scattering the transverse cross section is 
dominant in contrast to $\pi^+$ production where, for small $-t$, the longitudinal 
cross section is larger than the transverse one \ci{GK5,hermes-pi+,horn06}. 
This is a consequence of the fact that the kaon pole is much further away from the 
physical region than the pion one. Therefore, its contribution to the cross section 
is suppressed although its coupling constant and form factors are very similar to 
those of the pion. The relative suppression of the kaon cross sections is given by
\be
      \frac{d\sigma_L(K^+)}{d\sigma_L(\pi^+)} \sim \frac{(t-m_\pi^2)^2}{(t-m_K^2)^2}
\ee
at small $-t$. At small skewness and $t\simeq 0$ the suppression factor is approximately
\be
   \simeq (m_\pi/m_K)^4 \=0.63\cdot10^{-2}\,.
\ee

\begin{figure}
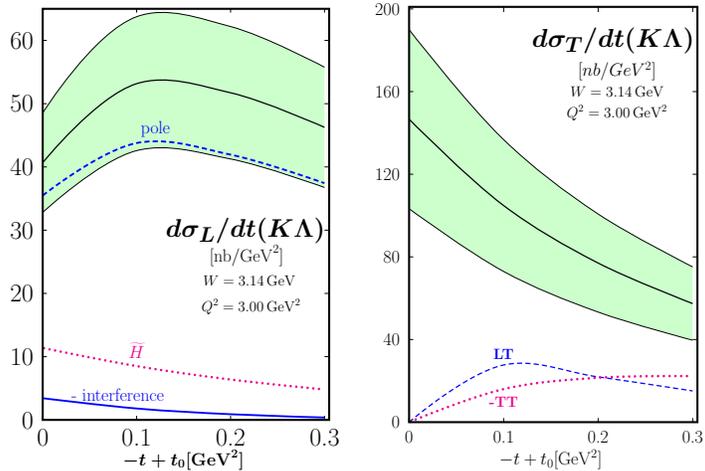

\centering
\includegraphics[width=0.36\tw]{fig-sigmaL-kaon.epsi}\hspace*{0.03\tw}
\includegraphics[width=0.36\tw]{fig-sigmaT-kaon.epsi}
\caption{The partial cross sections for $\gamma^* p\to K^+\Lambda$ 
versus $t-t_0$ at $W=3.14\,\gev$ and $Q^2=3.0\,\gev^2$. For the longitudinal cross section 
the kaon-pole, the $\widetilde{H}$ contributions as well as the interference between the 
pole and the GPD contributions are shown separately. This interference term as well as the 
transverse-transverse interference cross section are multiplied by $-1$. The shaded bands
indicate the parametric uncertainties of the theoretical predictions.}
\label{fig:kaon-t}
\end{figure}

The $t$-dependence of the longitudinal and transversal cross sections for the 
$K^+\Lambda$ channel are shown in Fig.\ \ref{fig:kaon-t} at kinematics typical for
the E12-09-11 experiment: $W=3.14\,\gev$ and $Q^2=3.00\,\gev^2$. The smaller contribution 
from the kaon pole also affects the shape of the longitudinal cross section. It starts
with a dip for forward scattering. Like for $\pi^+$ production the longitudinal cross
section is dominated by the pole contribution and the interference between the pole and
$\widetilde{H}$ contributions is negative. The contribution from $\tilde E$ is almost negligible. 
The transverse cross section is dominated by the contribution from $H_T$, the one from 
$\bar{E}_T$  only amounts to about $10\%$. The longitudinal-transverse and the 
transverse-transverse interference cross sections are also shown in Fig.\ \ref{fig:kaon-t}. 
They are markedly smaller in absolute value than the transverse cross section. 

There are also interesting polarization phenomena as for instance the correlation between the
helicities of the virtual photon and that of the target proton or the asymmetries measured with
a transversely polarized target. In general these polarizations are very similar in size to 
the case of $\pi^+$ productions \ci{GK5} but have occasionally the opposite sign.
 
One may also calculate the $K^+\Sigma^0$ channel. For a flavor symmetric sea only the 
$d$-valence quark GPDs contribute \ci{frankfurt99},
\be
K_{p\to\Sigma^0,i}\simeq - \frac1{\sqrt{2}}\,K_i^{d}\,,
\label{eq:GPD-S}
\ee
and the coupling constant $g_{KN\Sigma^0}$ is 3.5 \ci{compilation}. All other input parameters are 
the same as for the $K\Lambda$ channel. Since both the GPD as well as the coupling constant are 
much smaller than for the case of the $\Lambda$ the $K^+\Sigma^0$ cross sections are more than an 
order of magnitude smaller than the ones for the kaon-$\Lambda$ channel while the shapes of
the cross sections are similar in both cases.

\section{The kaon-induced exclusive Drell-Yan \\process}
\label{sec:drellyan}
Next the process $K^-(q,0) p(p,\nu)\to \gamma^*(q',\mu')\Lambda(p',\nu')$ 
(with $\gamma^*(q')\to l^-(k)l^+(k')$) will be investigated. It is treated in
full analogy to the case of a pion beam \ci{GK9}. Mandelstam $s=(p+q)^2$ as well as
the photon virtuality $Q'{}^2=(k+k')^2$ are considered to be large but the time-like
analogue of Bjorken-$x$ 
\be
\tau\=\frac{Q^{\prime 2}}{s-m^2}
\ee
is assumed to be small. The skewness is related to $\tau$ analogously
to Eq.\ \req{eq:skewness} by
\be
\xi\=\frac{\tau}{2-\tau}\,.
\ee
At large values of $\tau$ respective large values of $Q^{\prime 2}$ and fixed $s$ only large 
$-t$ contribute since $-t_0$ (see \req{eq:t0} which also holds for the Drell-Yan process) 
becomes large. For $-t$ larger than about $1\,\gev^2$ the GPDs are not well known. They are 
merely extrapolations from a region of smaller $-t$ \ci{GK5,GK6}. Assuming factorization 
the Drell-Yan amplitudes ${\cal M}_{\mu'\nu',0\nu}$ are expressed as convolutions of hard 
subprocess amplitudes and the same proton-$\Lambda$ transition GPDs as for electroproduction
of kaons. The subprocess amplitudes are the $\hat{s} - \hat{u}$ crossed electroproduction 
subprocess amplitudes \req{eq:subprocess}: 
\be
{\cal H}^{K^-\to\gamma^*}(\hat{s},\hat{u}) \= -{\cal H}^{\gamma^*\to K^+}(\hat{u},\hat{s})
\ee
where $\hat{s}$ and $\hat{u}$ denote the subprocess Mandelstam variables. The replacement
$\hat{s}\leftrightarrow \hat{u}$ is equivalent to replacing $\xi$ by $-\xi$ and taking the 
the complex conjugated amplitude. Thus, the calculation of the Drell-Yan process is analogous
to that one of electroproduction. The only difference is that in the pole residue \req{eq:residue} the 
time-like form factor of the kaon \req{eq:tl-FF} instead of the space-like one \req{eq:sl-FF} 
is to be used.

As for electroproduction there are four partial cross sections which are defined analogously
to electroproduction. The four-fold differential cross section for 
$K^-p\to l^-l^+\Lambda$ reads  \ci{GK9}
\ba
\frac{d\sigma}{dtdQ^{\prime 2}d\cos{\theta}d\phi} &=& \frac{3}{8\pi}\left\{
                         \sin^2{\theta}\,\frac{d\sigma_L}{dtdQ^{\prime 2}} 
          + \frac12 \big(1+\cos^2{\theta}\big)\,\frac{d\sigma_T}{dtdQ^{\prime 2}} \right. \nn\\
           &+& \left.\hspace*{-0.03\tw} \frac1{\sqrt{2}}\sin{(2\theta)}\cos{\phi}\, 
                       \frac{d\sigma_{LT}}{dtdQ^{\prime 2}}
            + \sin^2{\theta}\cos{(2\phi)}\, \frac{d\sigma_{TT}}{dtdQ^{\prime 2}}\right\}\,.
\ea
The azimuthal angle between the lepton and the hadron plane is denoted by  $\phi$
while $\theta$ is the decay angle in the rest frame of the virtual photon defined with respect 
to its direction in the center of mass frame.  

\begin{figure}
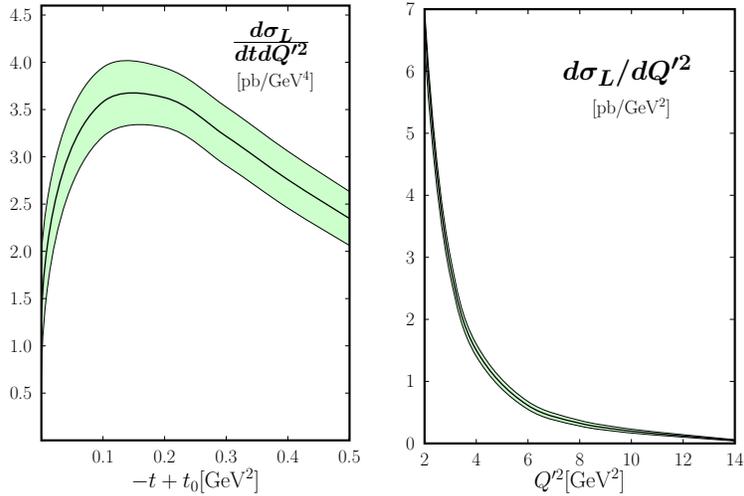

\centering
\includegraphics[width=0.385\tw]{fig-sigmaL-kaon-DY.epsi}
\hspace*{0.03\tw}
\includegraphics[width=0.37\tw]{fig-sigmaL-kaon-int-DY.epsi}
\caption{Left: The longitudinal cross section $d\sigma_L/dtdQ^{\prime 2}$ for 
$K^+ p\to l^+l^-\Lambda $ at $Q^{\prime 2}=4\,\gev^2$ versus $t$ (left)  and 
$d\sigma_L/dQ^{\prime 2}$ (right) versus $Q^{\prime 2}$. In both cases $s=30\,\gev^2$.
The shaded bands indicate the parametric uncertainties of the theoretical predictions.} 
\label{fig:sigmaL-kaon-DY}
\end{figure}
\begin{figure}
\centering
\includegraphics[width=0.35\tw]{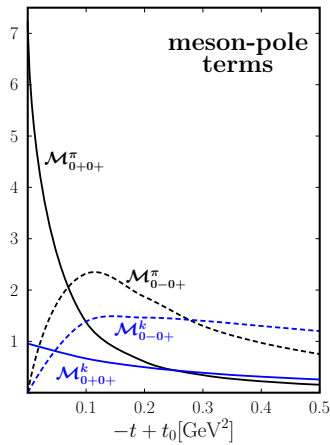}
\caption{The absolute values  of the pole contributions to the longitudinal helicity flip 
and non-flip amplitudes for pion and kaon induced Drell-Yan processes at $s=30\,\gev^2$ 
and $Q^{\prime 2}=4\,\gev^2$.} 
\label{fig:pole-terms}
\end{figure}

Results for the longitudinal cross section which is defined by (see also \ci{berger01})
\be
\frac{d\sigma_L}{dtdQ^{\prime 2}}\= \kappa\;\sum_{\nu'} |{\cal M}_{0\nu',0+}|^2
\ee
where the normalization factor reads
\be
\kappa \= \frac{\ale}{48\pi^2}\,\frac{\tau^2}{Q^{\prime 6}}\,,
\ee   
are shown in Fig.\ \ref{fig:sigmaL-kaon-DY} at $s=30\,\gev^2$ and at $Q^{\prime 2}=4\,\gev^2$ 
and integrated upon $t$. The longitudinal cross section reveals a deeper forward dip than 
in the space-like region. Such a dip is not seen in the case of the pion. The reason for 
this distinction is that the meson-pole term dominantly feeds the helicity-flip amplitude
which is of course suppressed for $t\to t_0$. However, in the case of the pion and for 
$Q^{\prime 2}/s \ll 1$, this effect is hidden  by a strong non-flip amplitude. For 
$Q^{\prime 2}/s \ll 1$ $t_0$ is small and, hence, $1/(t-m_{\pi}^2)$ becomes very large for 
$t\to t_0$. This can be seen from  Fig.\ \ref{fig:pole-terms} where the absolute values of 
the pole contributions to the longitudinal helicity flip and non-flip amplitudes are 
displayed at $s=30\,\gev^2$ and $Q^{\prime 2}=4\,\gev^2$. For larger values of $Q^{\prime 2}/s$ 
$-t_0$ becomes large and one is far away from the poles in both cases.  In contrast to the 
case of the pion \ci{chang,GK9,berger01} the results for the longitudinal cross section of 
the kaon-induced Drell-Yan process is not very different from a leading-twist calculation.

\begin{figure}
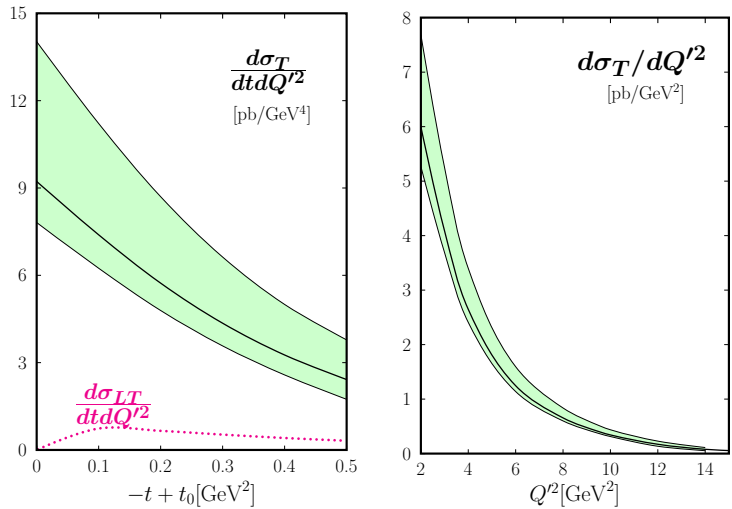

\centering
\includegraphics[width=0.38\tw]{fig-sigmaT-kaon-DY.epsi}
\hspace*{0.03\tw}
\includegraphics[width=0.36\tw]{fig-sigmaT-kaon-int-DY.epsi}
\caption{The transverse cross sections $d\sigma_T/dtdQ^{\prime 2}$ for $K^-p\to l^+l^-\Lambda$ 
at $Q^{\prime 2}=4\,\gev^2$ and $s=30\,\gev^2$ versus $t'$ (left)  and $d\sigma_T/dQ^{\prime 2}$ 
for the same process (right) versus $Q^{\prime 2}$. The shaded bands indicate the 
parametric uncertainties of the theoretical predictions.} 
\label{fig:sigmaT-kaon-DY}
\end{figure}

The transverse cross section, defined by \ci{GK9}
\be
\frac{d\sigma_T}{dtdQ^{\prime 2}}\=\kappa \sum_{\mu=\pm 1,\nu'}\;|{\cal M}_{\mu\nu',0\nu}|^2
\ee
is displayed in Fig.\ \ref{fig:sigmaT-kaon-DY}. It behaves similar to the one of the 
exclusive pion induced Drell-Yan process but is somewhat smaller. In contrast to the pion 
case where the longitudinal cross section is larger than the transverse one at small 
$Q^{\prime 2}/s$, both the kaon cross section are of similar magnitude. Again the different
size of the pole contribution is responsible for that. The by far dominant contribution
to the transverse cross section comes from the GPD $H_T$, the GPD $\bar{E}_T$ is almost 
negligible. Also shown in Fig.\ \ref{fig:sigmaT-kaon-DY} is the longitudinal-transverse 
interference cross section. It is very small. The transverse-transverse interference term 
is extremely small, about 0.1\, pb/GeV$^2$, and therefore not displayed in the figure. The 
definitions of the interference cross sections can be found in \ci{GK9}.

Despite the fairly large range of the photon virtuality considered here evolution is still 
a minor effect. For the longitudinal cross section this is so because the strong kaon pole 
term evaluated as a one-bose exchange, is not subject to evolution. The evolution of the 
transversity GPDs is a small effect as is discussed in Sect.\ \ref{subsec:GPD}. 
Therefore, the transverse cross section does not change much for $Q^{\prime 2}$ in the range 
of interest. For instance, it decreases by about $8\%$ at $Q^{\prime 2}=14\,\gev^2$ if the 
evolution of the transversity GPDs is taken into account as compared to the cross section
obtained without evolution. Of course, for very large $Q^{\prime 2}$ evolution matters - the 
transversity GPDs evolve to zero.

One has to be aware of the excitation of the charmonium states which appear as sharp, narrow spikes
in the Drell-Yan cross sections for photon virtualities near the masses of the $J/\Psi$ or the 
$\Psi(2S)$. The spread of the beam momentum will however widen the peaks considerably.
One may also detect the $J/\Psi$ directly, i.e.\ measure the cross section of the process
$K^-p\to J/\Psi n$. The above calculation of the exclusive Drell-Yan process allows
for an estimate of the electromagnetic contribution to this cross section. Such an estimate
is beyond the scope of the present work.

\section{Summary}
\label{sec:summary}
In this article the hard exclusive electroproduction of kaons as well as the 
crossed process, the kaon-induced exclusive Drell-Yan process, have been investigated
within the handbag approach. As for the analogous processes involving pions the 
transversity GPDs play an important role. Their use goes along with a twist-3
meson wave function which is applied in Wandzura-Wilczek approximation. 
It would be interesting to go beyond this approximation and to include both the two-
and three-body twist-3 contributions. In wide-angle photoproduction of pions at least 
these contributions play an important role \ci{passek18}. In contrast to previous
studies of hard processes involving pseudoscalar mesons the evolution of the transversity
GPDs is taken into account. It turns out, however, that this evolution effect is
small in the range of photon virtualities of interest currently.

Predictions for the various cross sections are given and compared to the available data.
Fair agreement with experiment \ci{mohring,coman,carmignotto} is observed.
At present there is no significant signal of contributions from sea quarks neither to 
$\pi^+$ nor to $K^+$ production. This situation may change with the advent of more and
better data on these processes as, for instance, can be expected from experiments
performed at the upgraded JLab. Nothing is known as yet on the GPDs $\tilde E$, $H_T$
and $\bar{E}_T$ for sea quarks. Only the forward limit of $\widetilde H$
for sea quarks, i.e.\ the combination 
$-2(2\widetilde{H}^{\bar{u}}-\widetilde{H}^{\bar{d}}-\widetilde{H}^s)/\sqrt{6}$ can be
estimated. According to the DSSV polarized parton densities \ci{DSSV09}, it is small and
has a zero at $x\simeq 0.18$. Combined with a Regge-like profile function as in
\req{eq:profile} one finds that the contribution from this GPD to the longitudinal
electroproduction cross section is small well within the uncertainties of the theoretical
predictions.

\begin{appendix}
\section{Solving the  evolution equation of the transversity GPDs}
\label{sec:app}
The evolution of the quark transversity GPDs with the scale, $\mu^2$, is controled by the
integro-differential equation 
\be
\frac{d}{d\ln{\mu^2}} K_T(x,\xi,t,\mu^2)\=-\frac12\;\int_{-1}^1dy P(x,y,\xi)\,K_T(y,\xi,t,\mu^2)
\label{eq:int-diff}
\ee
where, for quarks, the evolution kernel reads \ci{bel-mul,bel-rad}
\ba
P(x,y,\xi)&=& \frac{\als}{2\pi} C_F \left\{ \Big[\frac{\xi+x}{x-y}\;
                    \frac{\Theta(\xi+x)\Theta(y-x)-\Theta(-\xi-x)\Theta(x-y)}{\xi+y}\Big]_+
                       \right.\nn\\
          &&\left.\hspace*{0.085\tw}+ \Big[\;\frac{\xi-x}{y-x}\;
                   \frac{\Theta(\xi-x)\Theta(x-y)-\Theta(x-\xi)\Theta(y-x)}{\xi-y}\;\Big]_+
                       \right.\nn\\
          &&\left.\hspace*{0.085\tw}+ \frac12\;\delta(x-y)  \right\}\,.
\ea
The symbol $[\ldots]_+$ denotes the usual plus prescription and $C_F=4/3$.      
It is important to note that there is no mixing with the corresponding gluon GPDs, quark 
and gluon transversity GPDs evolve independently from each other. Following Vinnikov 
\ci{vinnikov} logarithmic grids for $x$ and the scale $\mu^2$ are introduced:
\ba
x_i&=&\phantom{-}\delta \big(1- e^{-\gamma(i-2n)}\big)   \hspace*{0.1\tw} {\rm for}\; 
                                              -1\leq x\leq 0\,\quad\;\, 0\leq i\leq 2n \nn\\
x_i&=&-\delta \big(1- e^{\gamma(i-2n)}\;\big)\;   \hspace*{0.1\tw} {\rm for}\;\quad\; 
                                      0 < x\leq 1\, \quad 2n < i\leq 4n\,.
\ea
The variables $\delta$ and $\gamma$ are defined by
\be
\gamma\=\frac1{n}\ln{\frac{1-\xi}{\xi}}\,, \qquad \delta\=\frac{\xi^2}{1-2\xi}
\ee
and, for the evolution from $\mu_0$ to $\mu_1$,
\be
u_j\=u_0+jd\,  \qquad d\=\frac{u_1-u_0}{m} \qquad 0\leq j\leq m-1
\ee
where $u=\ln{\mu^2}$. With the help of the $x$-grid the integral in \req{eq:int-diff}
is replaced by a sum using Simpson's rule. This transforms the integro-differential 
equation into a system of coupled differential equation. This system is subsequently solved 
with the help of the 4th-order Runge-Kutta procedure. The numerical method is quite 
stable and for $n=20$ and $m=1(2)$ (for $\mu^2\leq 10 (20)\,\gev^2$) good results
have been obtained. A numerical code for calculating the evolution of the quark 
transversity GPDs can be obtained from the author on request.

\end{appendix}


\end{document}